\documentclass[english,aps,onecolumn,superscriptaddress,showpacs]{revtex4}
\usepackage[T1]{fontenc}
\usepackage[latin9]{inputenc}
\usepackage{textcomp}
\usepackage{amsmath}
\makeatletter

\usepackage{epsfig}
\usepackage{changebar}
\usepackage{epsfig}
\usepackage{subfigure}
\usepackage{rotating}
\usepackage{amsfonts}
\newlength {\oldtextheight}
\newlength {\oldheadsep}
\newcommand{\beq}{\begin{equation}}
\newcommand{\eeq}{\end{equation}}

\usepackage{color}

\newcommand{\bea}{\begin{eqnarray}}
\newcommand{\eea}{\end{eqnarray}}

\psfigdriver{dvips}

\usepackage{babel}
\makeatother

\begin{document}

\title{Entropic algorithms and the lid method as exploration tools for complex landscapes}
\author{Daniele Barettin}
 \affiliation{Mads
Clausen  Institute for Product Innovation, University of Southern
Denmark, 6400 S\o nderborg, Denmark}
\author{Paolo Sibani}
\email[]{paolo.sibani@ifk.sdu.dk}
\affiliation{Institut for Fysik og Kemi, SDU, DK5230 Odense M, Denmark}

\date{\today}

\begin{abstract}  
Monte Carlo algorithms such as the   Wang-Landau 
algorithm  and similar  `entropic' methods are able to 
accurately sample the density of states of model systems 
 and thereby give access to thermal equilibrium properties at any temperature.
Thermal equilibrium is however  not achievable at low temperatures in  glassy systems.
Such systems  are  characterized by a multitude of metastable configurations
pictorially referred to as `valleys' of  an energy landscape. 
Geometrical properties  of the landscape,
e.g. the local density of states describing the distribution in energy
of the states belonging to a single valley, are key to understand   the 
dynamical properties of such  systems.
In this paper we combine the lid algorithm, a tool for landscape 
exploration previously applied to a range of models, with the Wang-Swendsen 
algorithm. To test this improved  exploration tool, 
 we consider a paradigmatic complex system, the
Edwards-Andersom model in two and three spatial dimension.
We find a striking difference between the energy dependence 
of the local density of states in the two cases: flat in the first case, and nearly exponential 
in the second. The lid dependence of the data is analyzed to estimate the form of the
global density of states.  
\end{abstract}
\maketitle
\section{Introduction}
\label{introduction}  
Energy landscapes of model systems have been studied extensively using  a variety of numerical 
methods specialized for different purposes \cite{Wales03}. 
E.g., using repeated  thermal quenches, each quench  leading to the local energy
minimum configuration or \emph{inherent state} \cite{Stillinger83} which lies `below' the current 
state of a Monte Carlo (MC) simulation, the landscape is partitioned  into the catchment basins 
belonging  to the different inherent states. Identifying the  important connections 
between these basins, i.e., typically, the `mountain passes' of lowest energy, one can  then produce a 
coarse-grained version of the landscape and use it to assess both dynamical and equilibrium properties. 
This  approach encounters problems  in models of glassy systems, 
which feature  a quasi-continuum of metastable configurations, each configuration parameterized by
an energy barrier gauging  its  thermal stability  and its life-time under iso-thermal conditions.

If  equilibrium thermal properties are of interest, `entropic' or `flat histogram' methods 
\cite{Ferrenberg88,Wang01,Lee93} are powerful and generally applicable tools. These methods avoid the 
trapping in local energy minima which plagues the standard Metropolis algorithm and produce 
the global density of states (GDOS) as a function of 
the energy. From there,  any  thermal equilibrium property  of interest can be obtained.
Since, however, thermal equilibrium is not experimentally achievable in glassy systems
at low temperatures, equilibrium properties of pertinent  models have mainly academic interest. 

Local geometrical features  of the energy landscape, as extracted by  the lid method have a  
direct bearing on the dynamical properties of glassy model systems: a fictitious  and impenetrable 
energy barrier, called `lid', is  introduced and the energy distribution of all the micro-states which 
can be reached starting from a given inherent state without ever crossing the lid energy is determined. 
The distribution  is called Local Density of States (LDOS) and gives access to the local equilibrium 
properties of the (fictitious) valley jointly defined by an inherent state and a lid, just like the 
density of states provides corresponding global information. Furthermore, the thermal stability of a 
valley at a given temperature is encoded in the LDOS. 
 
Finding the LDOS can be numerically  challenging:  Metropolis  sampling at low temperatures is 
hampered by  the large number of local minima, while 
at high temperature, trajectories linger in many cases 
just below the lid energy due to the great number of states available there.
In both cases, the result is poor sampling.
Exhaustive enumeration was utilized  in Refs. \cite{Sibani93,Sibani94,Schon98,Schon00,Sibani99}  
to avoid these problems.  That  procedure is both fast and exact, but puts very  high
demands  on memory availability  and  quickly runs out of space when the system size and/or the 
energy range is large.

In this paper we show how the LDOS can be efficiently estimated  
by combining the lid method with the flat 
histogram method of Wang and Landau \cite{Wang01}.
From an algorithmic point of view, imposing a lid restricts the search space. While this hardly changes the 
overall convergence properties of the histogram method, the  lid makes it possible
to obtain a far more  accurate description of the density of states in the low  energy region close to
the ground state of the model investigated.   As shown below, this accuracy is not guaranteed
by unrestricted entropic method, since  the  tiny fraction of all the available configurations
present near the ground state can  be  extremely hard to sample.
Compared to the lid method in combination with exhaustive search, the present approach is 
far more powerful: larger systems can be investigated over a much larger range of energies.

In this paper the   algorithm   is applied to  the Edwards-Anderson (EA)\cite{Edwards75} spin glass model, 
a paradigmatic glassy systems featuring quenched disordered
interactions. We find that  the energy dependences of the  LDOS   in  two dimensions (2D)  and three dimensions (3D)
 differ considerably, and that the 3D results 
concur with  important experimental features of real spin glasses.
\section{Method  and model}
\label{model} 
Since  key  properties of experimental spin glasses \cite{Edwards75}
are reproduced by Monte Carlo simulations of the EA model, the latter can be considered 
 a \emph{bona fide}  complex system in its own right \cite{Picco,Batt}. For completeness, we briefly
 describe the model and the relevant features of its energy landscape before turning to how
 entropic sampling and the lid method are jointly applied to the EA model.
 \subsection{The Edwards-Anderson model and the lid method}
In the  EA model  Ising spins, $S_i= \pm 1$, are placed on a cubic  lattice
and interact with their    nearest-neighbors  through quenched random couplings.
A system of $N$ Ising spin in configuration $x$  
has  an energy $E(x)$ given by  
\bea
E(x) =-\frac{1}{2}\sum_{ij}J_{ij}S_i^x S_j^x - H \sum_{i}S_i^x,
\label{energy_def}
\eea
where  the coupling matrix $\mathbf J$  is symmetric, with  elements $J_{ij} $ 
vanishing  unless the 
sites $i$  and $j$ are neighbors on the lattice.
For $i<j$, the non zero couplings are  random independent variables, drawn in our case
 from  a  Gaussian distribution, with zero average and unit variance.
The last  term in Eq.\eqref{energy_def} describes the interaction with an external magnetic field $H$. 

In 3D,  at $H=0$ and for temperatures below a critical temperature $T_c\approx 0.95$, (see Ref. 
\cite{Katzgraber06} for a chronologically
ordered list of different $T_c$ estimates) the model  is in its  spin-glass phase, which is characterized by a 
non-zero value of the spin-glass order parameter \cite{Fischer91, Parisi}. 

At low energies, the energy landscape of the EA model  features  a multitude  of nested metastable 
`valleys' i.e., regions of configuration space which support a metastable equilibrium-like  
probability distribution \cite{Sibani94}. A valley  jointly defined by an inherent state of energy
$E_{\rm min}$ and a  lid energy $E_{\rm lid} >E_{\rm min} $ comprises 
all configurations connected to the inherent states by paths
(i.e.  series of single spin flips)   never  crossing the lid energy.
For each valley exhaustive enumeration of all states below the lid
is implemented up to the  lid value
which opens a connection to a new  inherent state of lower energy \cite{Sibani94}.  
For rather small systems, the LDOS thus obtained seems
 to grow in a near exponential fashion (see also \cite{Klotz98a}). The lid dependence of the 
 rate of growth has not been clarified, nor is it clear how the property extends to larger systems.

\subsection{The entropic sampling algorithm} 
Entropic sampling   is a Monte-Carlo technique invented by Lee \cite{Lee93} where transitions
are controlled by   (the current estimate of) the density of states $\varrho(E)$.
Rather than sampling with the usual Boltzmann weight $e^{- \frac{E}{T}}$, the entropic method samples with a probability 
$\propto \frac{1}{\varrho(E)}$. 
Unlike the energy,  the density of   states $\varrho$ is not known beforehand and must hence  be
calculated iteratively during the simulation.   Starting with a (poor) guess 
$\varrho_1(E)=$ constant, we divide the energy axis into a certain number of bins,
calculate a histogram $h_1(E)$ of the energies of the states visited and normalize it to one.  The probability  $h_1(E)$
to visit an energy bin $E$ is proportional to the number 
of states $\varrho_1(E)$  multiplied by the probability     $1/\varrho_1(E)$ with which we sampled, i.e. :
\bea
\varrho(E) \propto \varrho_1(E)h_1(E).
\eea 
We  use the above to iteratively define 
\bea
\varrho_{n+1}(E)=\varrho_n(E)h_n(E)
\label{iteration}.
\eea 
which specifies  the algorithm in terms of successive approximations to the density of states. 
In terms of the micro-canonical entropy 
\bea
S(E)=\log \varrho(E),
\eea
the algorithm reads
\bea
S_{n+1}(E)=S_n(E)+\log h_n (E),
\eea
from which is clear that  the  probability of sampling a particular state is proportional to $e^{-S(E)}$, 
and that convergence implies $h_n (E)  \rightarrow 1$ for $n\rightarrow \infty$. 
For a general discussion of entropic algorithms, we refer the reader  to the book by Newman and Barkema \cite{Newman99b}.

The entropic sampling algorithm applied in conjunction  with the lid method yields the LDOS previously mentioned.
The first step     is to identify a local energy minimum using a 
  Monte Carlo algorithm running at  constant temperature.
The energy of this minimum is taken as the zero of the energy axis, 
and as the lower edge of the first bin in the energy histogram to be constructed. 
The standard entropic algorithm is modified by adding a rejection criterion:
every spin configuration with an energy greater then the lid is 
rejected \textit{a priori}. The lid value will therefore  be 
the upper edge of the last bin in the energy histogram.
 Finally,  a `bail-out' option is included: 
whenever a configuration of negative energy is found, a new lower energy minimum is identified
and the whole process starts afresh with that new minimum used as starting point.
 \section{Results} 
Below,  we present results for the LDOS, $\varrho(\epsilon)$,  for the  EA model on square
and cubic lattices of linear size $L=30$ and $L=8$, respectively. The energy is in all cases
scaled by the number of spins $N_{\rm spin}$, i.e., $\epsilon = E/N_{\rm spin}$.
The average  of the  
 lowest energy values encountered in each of the $N_{\rm sample}$ different runs is denoted 
$\epsilon_{\rm min}$.
We consider four different 
valleys  defined,  in units of energy per spin, by the lid values $\lambda= 0.1,0.2,0.4$ and $0.8$.
The number of bins in the histogram is $N_{h}=20$ for $\lambda=0.1, 0.2$, and $N_{h}=40$ and
 $N_{h}=80$ for $\lambda=0.4$ and 
$\lambda =0.8$, respectively. We also consider for comparison the Global Density of States (GDOS) 
obtained by the entropic algorithm with no lid restrictions imposed.
Unless otherwise specified,  the sum of the density of state over all
available energy bins   is normalized to one.

Two types of error may affect the calculations: the first type  is  lack of convergence due 
to an insufficient number of iterations being performed. As our results show, the unrestricted
entropic algorithm definitely suffers from this problem near the endpoints of the energy range.
The issue is therefore  the convergence of the algorithm  when a lid is imposed.
All  results presented being   averages
over  $N_{\rm sample}=100$ different realizations of the couplings, the second error type
is the statistical error on   these averages.

In  the entropic algorithm,    
a large number of iterations, i.e. reaching a large $n$ value in Eq.\eqref{iteration},
is more important  than obtaining  an accurate  
 histogram at any particular iteration as discussed in Ref.~\cite{Newman99b}. 
A relatively high number of iterations $N_{\rm iter}=40000$, and a relatively low number 
of MC sweeps at each iteration, $N_{\rm sweep}=20$ is therefore chosen. The weights $h_i(E)$ of the first 
50 iterations are disregarded  and these iterations only provide   
a less naive starting guess for the density than the uniform one.

\begin{figure}
 $
\begin{array}{cc}
\rotatebox{-90}{
\includegraphics[width=0.33\linewidth]{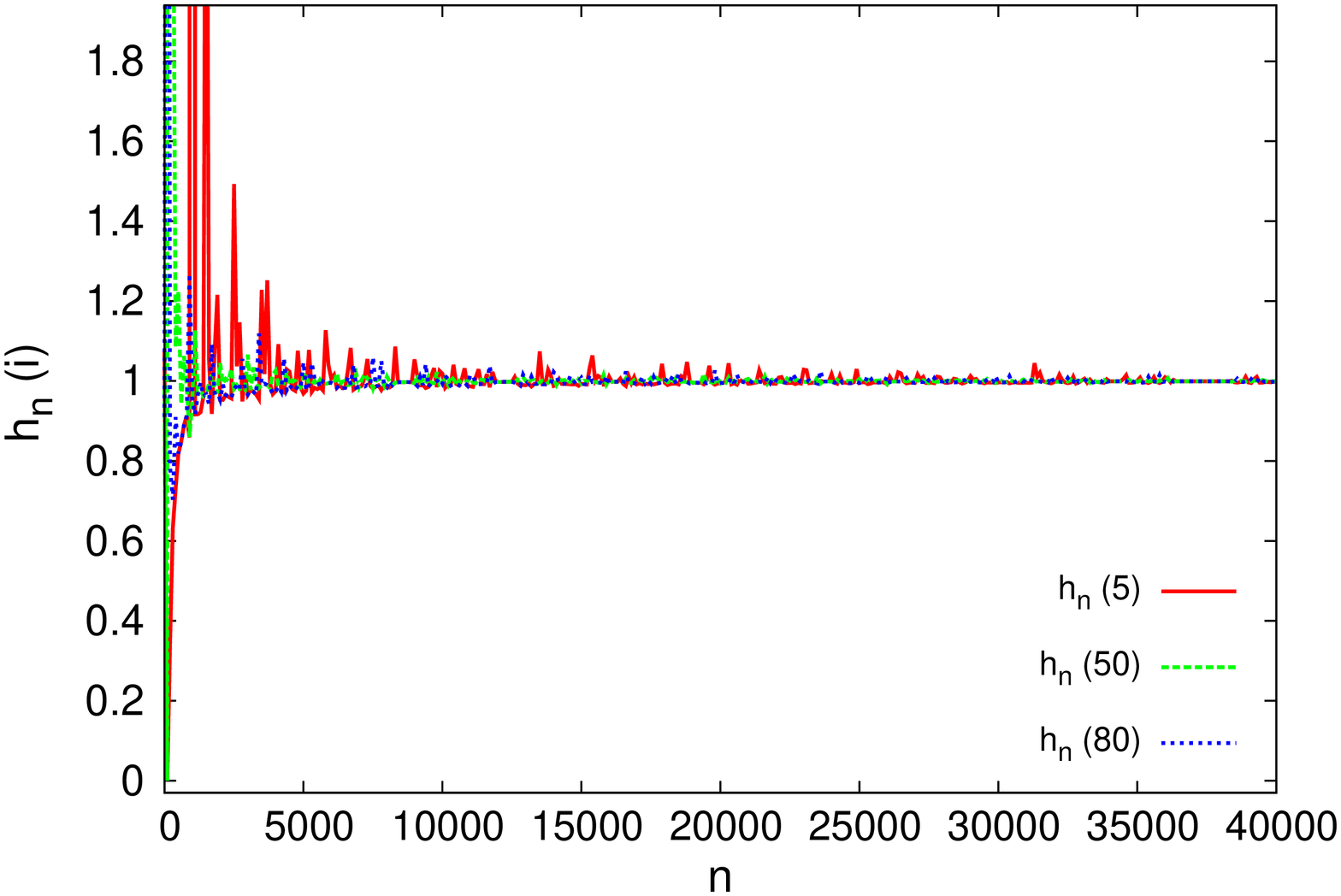} } & 
\rotatebox{-90}{\includegraphics[width=0.33\linewidth]{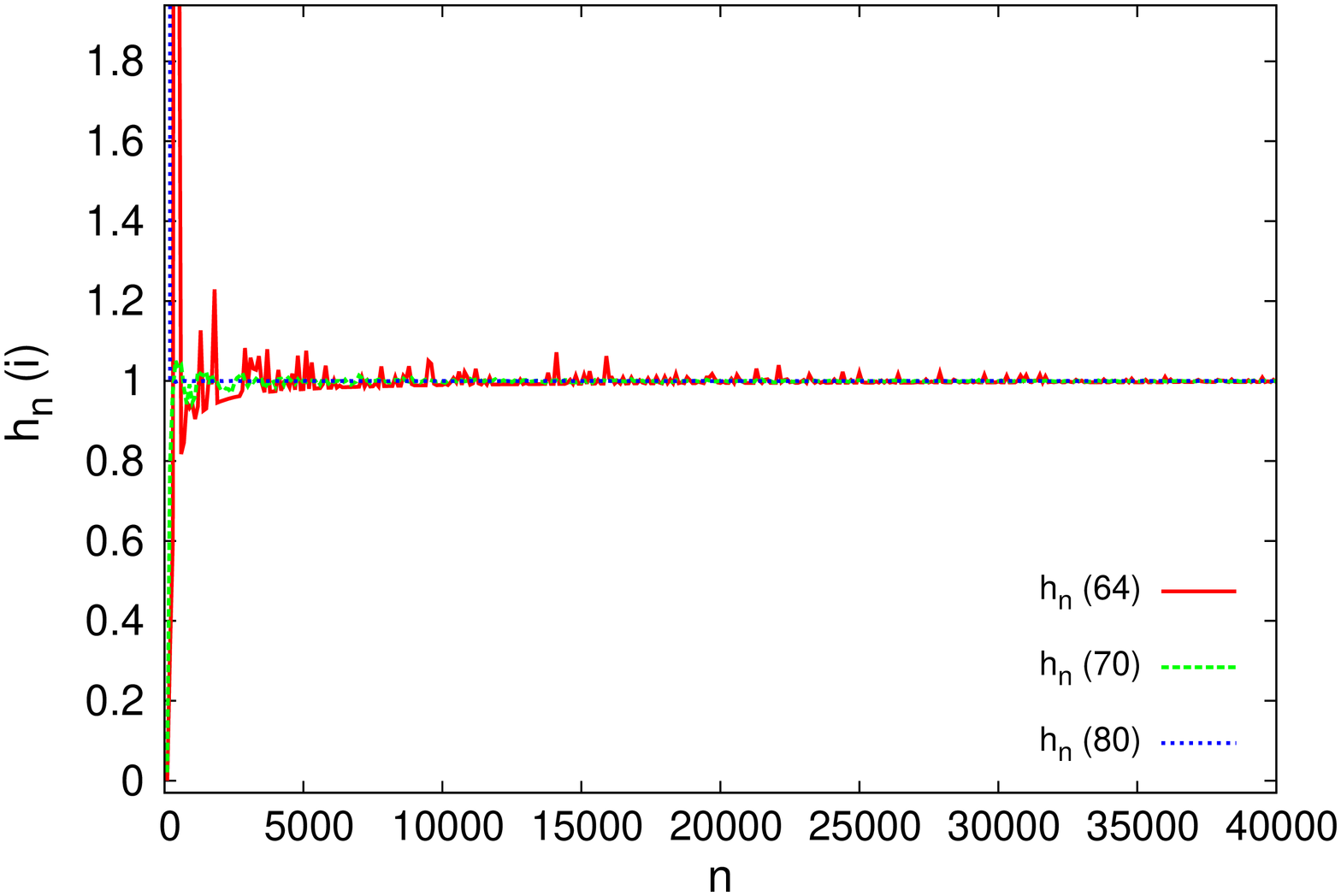}}
\end{array}
$ 
\caption{(Color online)
Each curve depicts the  ratio of two values of the LDOS taken $100$ iterations apart.
The simulations are all performed at lid value $0.8$.
Left: 2D system.
The three curves shown correspond to values $5, 50$ and $80$ of  the energy  bin index,  $1\le i \le 80$.
 Right: 3D system. The  energy bin  values for the three curves are here 
  $64,70$ and $80$.}
\label{convergence} 
\end{figure}

The  convergence of the algorithm (with a lid imposed) is investigated  using  the deviation from unity of the ratio 
between  approximations of  the density of states taken $m$ iteration steps apart,  
$h_n(i) =\varrho_{n+m}(i)/ \varrho_n(i)$, 
where $1\le i \le 80$ is  the energy bin index, $n$ is the iteration number and where $m$
is  for convenience taken to be much larger than one. 
In Fig. \eqref{convergence} we show the results of
calculations carried out, in both 2D
and 3D for  a single realization of the bonds,  using  $m=100$, three values of $i$ and    $\lambda = 0.8$.
 The choice  $\lambda=0.8$   covers the case 
 where   good  convergence
is hardest   to achieve  as the size of the configuration space to be explored
is  largest.
The three bin index values   used show  typical behavior. I.e. very similar curves
are obtained for other values of $i$. There is, however, a \emph{caveat}: since in 3D and for $\lambda =  0.8$
only a small fraction of the states lies 
near the lower edge of the energy spectrum,  the statistical sampling
is insufficient  in this region (a problem easily fixed by choosing a smaller
$\lambda$). The bins for the 3D case are accordingly  all chosen to  lie 
in the higher part of the energy interval. 
In general, our  results show that, after an oscillatory transient,  the algorithm reaches
a good convergence  in the $40 000$ iterations we have used throughout this work.

We now turn to the error on the  LDOS averaged over many different samples.
Since the  sample to sample fluctuations of $\varrho (\epsilon)$ are rather small, 
the statistical  errors in the graphs shown are  utterly negligible. To obtain a more quantitative 
assessment,  the   error bar $\sigma(\epsilon)$ on the  data presented is  calculated 
as the standard deviation of the average  LDOS, estimated over the  $N_{\rm sample}$ independent
simulations and divided by $\surd N_{\rm sample}$. The relative statistical error is then given  
for each value of the energy as  the ratio of $\sigma(\epsilon)/\varrho (\epsilon)$. The above procedure was carried out
in both 2D and 3D, and in each case for the two lid values at the opposite ends of the range investigated, i.e. 
$\lambda = 0.1$ and $\lambda = 0.8$. 
The largest values of the relative error 
observed through the energy range  provide relevant   bounds. In 2D, these 
bounds are   $4$\% and $7$\% for $\lambda = 0.1$ and $\lambda = 0.8$,  respectively.
The corresponding values for the 3D system are $4$\% and $10$\%. These bounds   all 
stem from  the lowest 
energy  bin, and hence belong to the smallest  value of the LDOS. Since the relative  error decreases
very rapidly  with energy,  they give a rather  pessimistic view of the uncertainty on our data. 

In  2D, the entropic algorithm repeatedly encounters states of energy lower than the 
energy of the  `current' lowest state,   meaning that the search is repeatedly abandoned and restarted
from  the new lowest state. This behavior is connected to the form of the LDOS, which is,
as we shall see, almost flat, except for the lowest energies. By contrast, the LDOS increases very rapidly with  energy in 3D.
This prevents the algorithm  from fully exploring the relatively few 
states located at energies near the bottom of the current valley.
Correspondingly, states of energy lower than the bottom
 of the current valley may go unnoticed and the 
 search is only restarted in a few cases.
For the same reason, an  unconstrained entropic algorithm  is incapable of sampling a  large 
 fraction of states at the low (and high) end of the energy spectrum. Using data obtained at different lids,
  it is  however possible to 
 patch together an accurate  density of states  spanning  approximately $40$ orders of magnitude.
\begin{figure}
 $
\begin{array}{cc}
\rotatebox{-90}{
\includegraphics[width=0.33\linewidth]{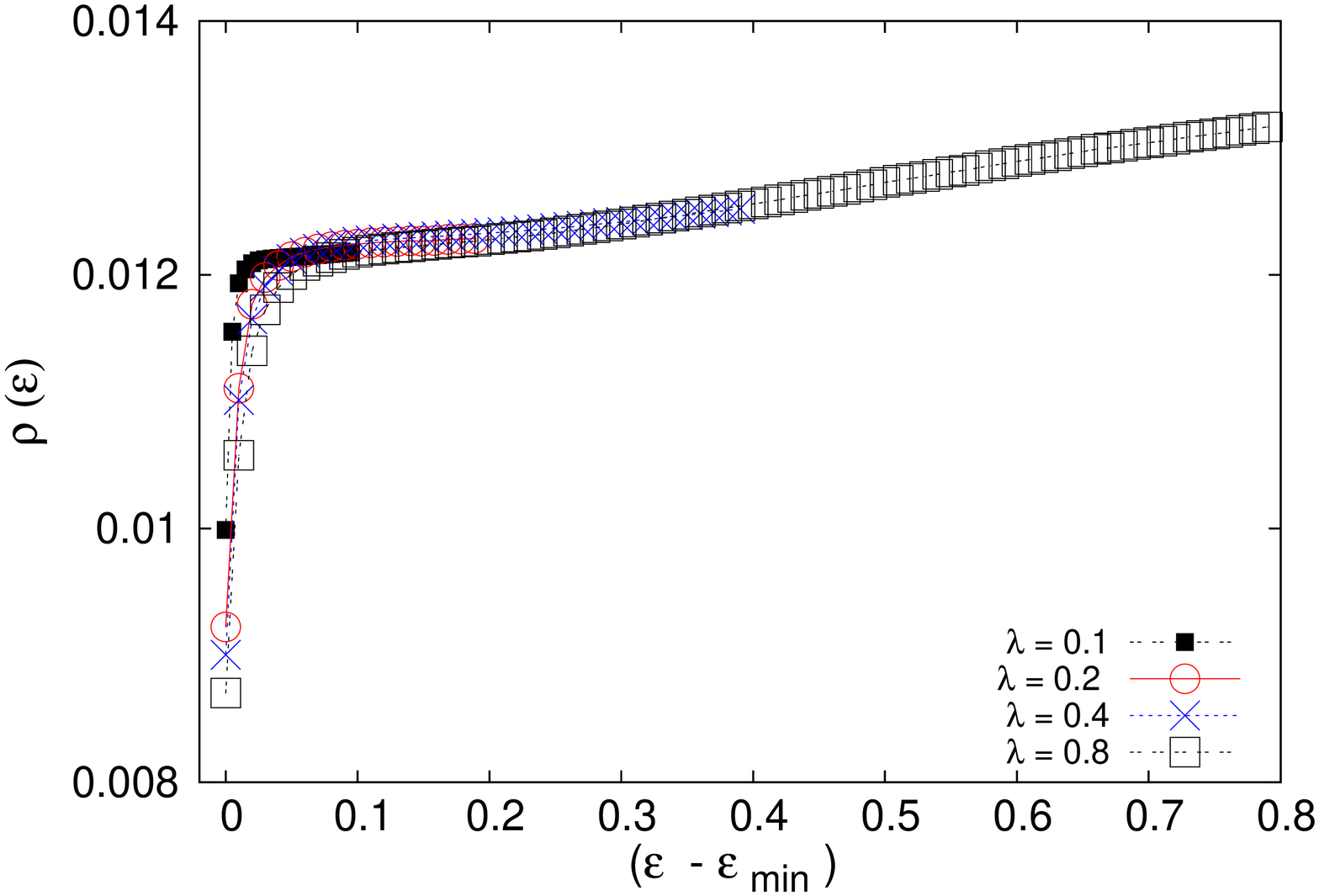} } &
\rotatebox{-90}{\includegraphics[width=0.33\linewidth]{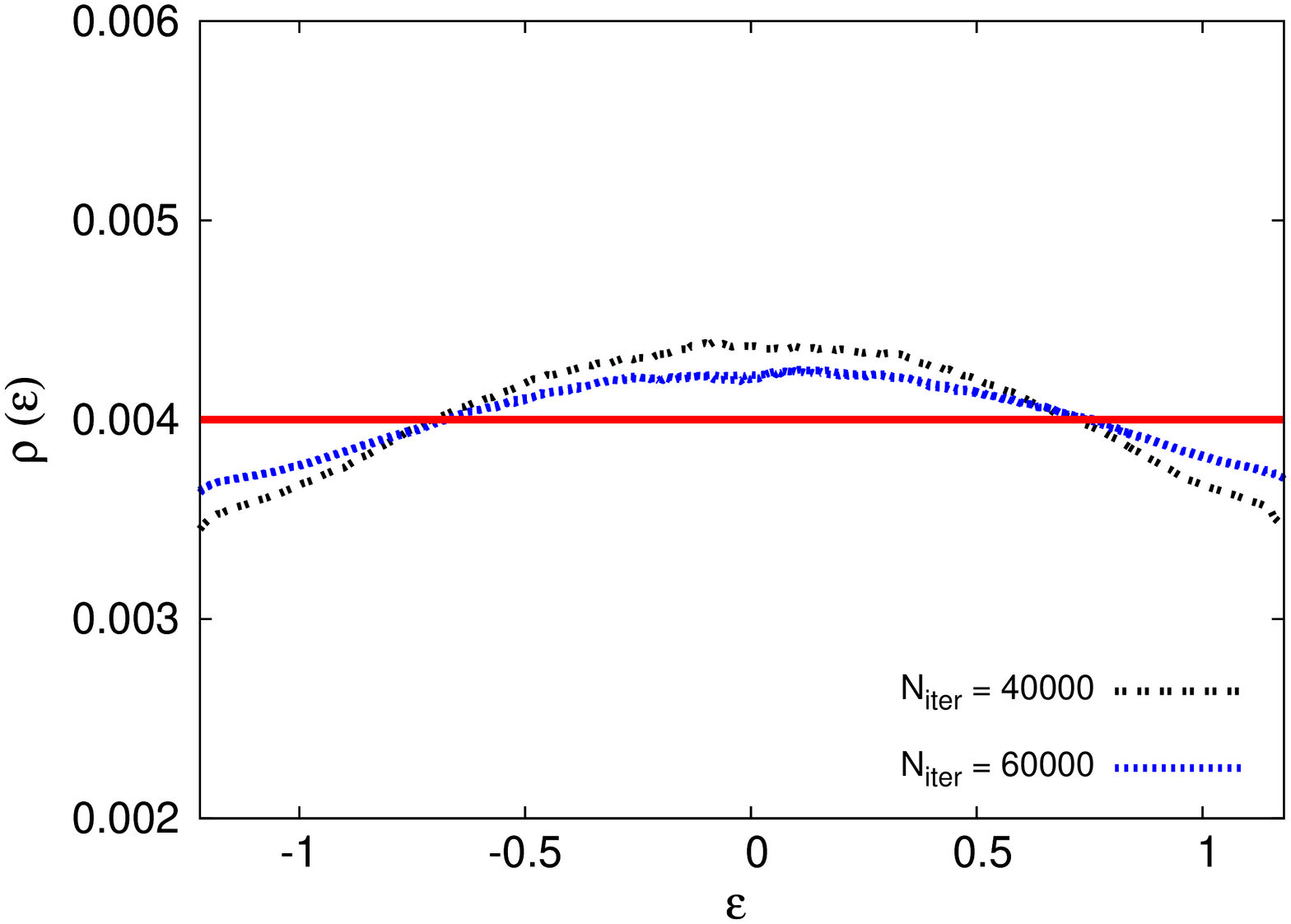}}
\end{array}
$ 
\caption{(Color online)
Left: LDOS for a 2D lattice with $L=30$ for the four different lids $\lambda=0.1,0.2,0.4,0.8$. 
Right: The GDOS $\varrho (\epsilon)$ for two
different values of the iteration number  $N_{\rm iter}$.  All data are for  a 2D lattice with $L=30$.}
\label{2D} 
\end{figure}  
\subsection{Two spatial dimensions}
Our first set of results pertains to a 2D square lattice with $L=30$. In the left  panel of Fig.\ref{2D},
 the 
LDOS is    plotted  on a linear  vertical scale versus  $(\epsilon-\epsilon_{\rm min})$ for  four different lid values.
The three curves corresponding to the lower lids have been vertically shifted in order 
to make them coincide with the fourth curve.
Near the ground state, the LDOS increases very rapidly with energy.  At higher energies the growth
tapers off and is nearly linear in the rest of the  range explored. 
 The GDOS depicted in the right panel of the figure  are  obtained
by running the entropic algorithm without any lid constraint. 
Two data sets are shown,  obtained using    $N_h=250$ in both cases and a   number of iterations
 equal to $N_{\rm iter}=40000$ and $N_{\rm iter}=60000$, respectively. For these two curves the data are averaged over $10$
samples. 
For comparison, a horizontal  line is  drawn  as 
a guide to the eye.
The    curvature  of  the GDOS is seen to  decrease slightly 
as the number of iterations increases, indicating a certain (and expected)  lack of convergence 
of the unrestricted algorithm. Nevertheless, the relatively modest range of $\varrho$
guarantees a passable   sampling of   the full range  of energies
available, including  the region near  the ground state energy of the model.
As we shall see, the same situation is not  achieved by the unrestricted
algorithm in the 3D case.
 
A simplified  cartoon picture   of the energy  landscape of the 2D EA spin glass 
is as follows:  the landscape contains a series of   valleys, all similar with respect to 
the internal distribution in energy of their respective configurations. 
Configurations belonging to different valleys have   non-overlapping  energies, 
i.e., the lowest lying state  of one  valley lies above  the
upper rim of the valley located just below it.  The valleys all have 
 a slowly growing  LDOS, and since 
the global density of states at 
a given energy  mainly counts  states belonging to  the single valley located near  that energy,
the GDOS is likewise slowly growing.
Note that if we disregard the initial transient,
$S(E) = \ln(\varrho(E)) \propto \ln(E) $, whence the average energy is simply proportional to the temperature.
This excludes any   thermal instability  of the kind shown below to be present in 3D.
\subsection{Three spatial dimensions} 
\begin{figure}
 $
\begin{array}{cc}
\rotatebox{-90}{
\includegraphics[width=0.33\linewidth]{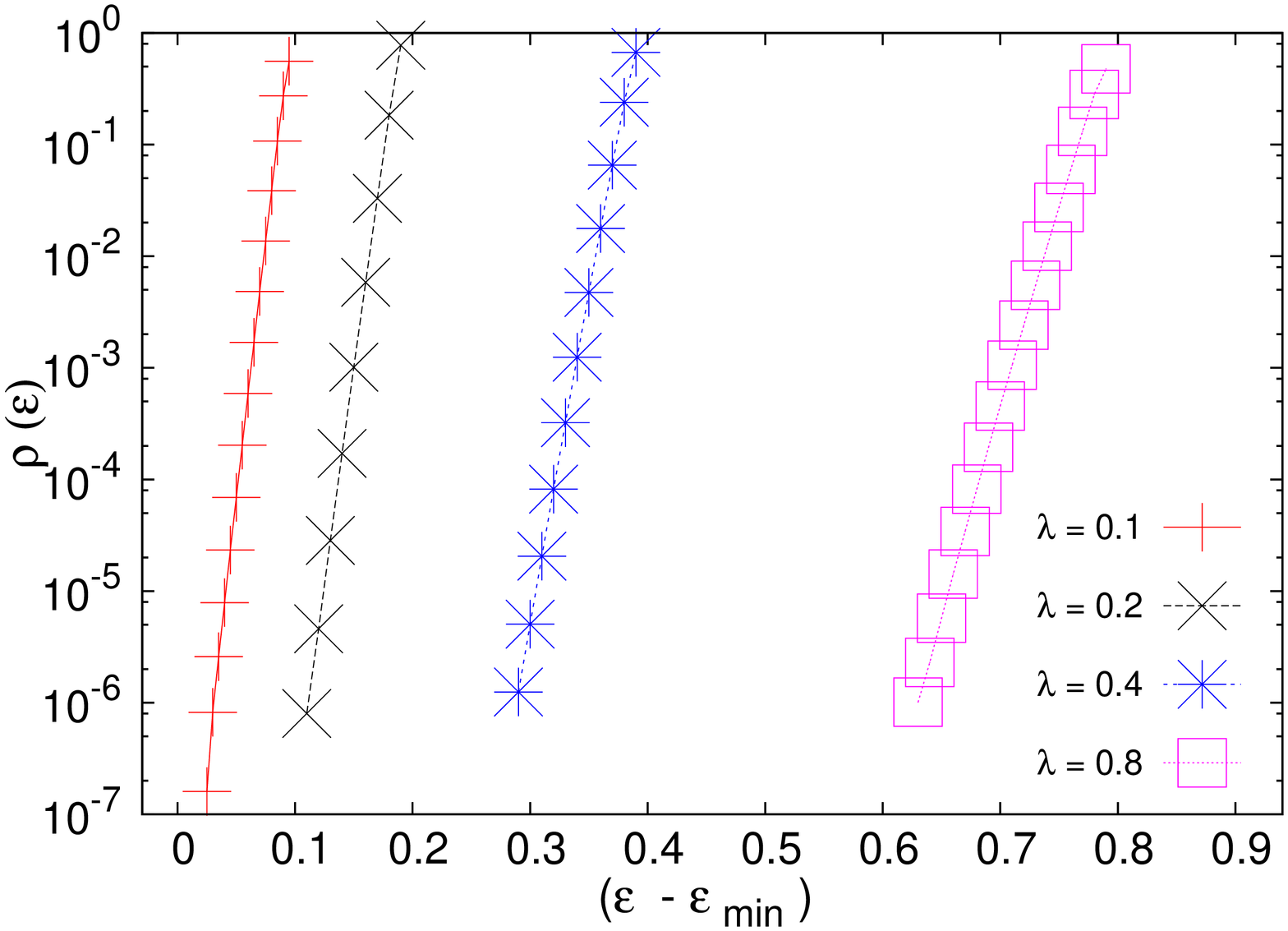} } & 
\rotatebox{-90}{\includegraphics[width=0.33\linewidth]{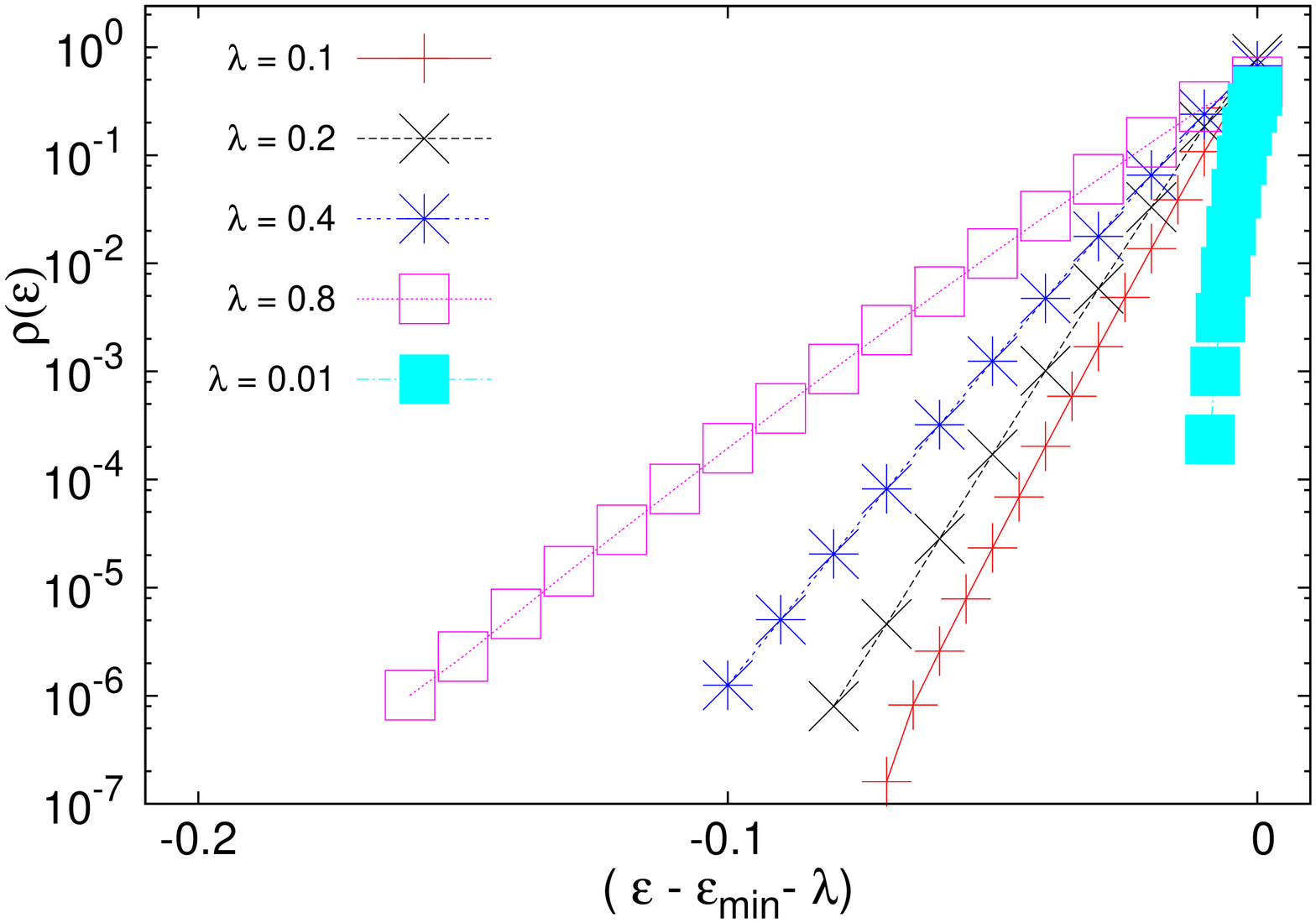}}
\end{array}
$ 
\caption{(Color online)
Left:  $\varrho (\epsilon)$,  for the four different lids
 $\lambda=0.1,0.2,0.4$ and $0.8$. Right:
The same four quantities together with  an additional LDOS obtained for $\lambda=0.01$  
are   plotted on  shifted  horizontal scales ending in each case at the lid value. The  
  slope of the curves is clearly seen to decrease with increasing lid value.
All data pertain to  a 3D lattice with $L=8$.}
\label{3D1} 
\end{figure}

 Figure~\ref{3D1} shows  the LDOS obtained for  a 3D cubic lattice with linear size $L=8$. 
 In the left panel, data obtained  for  four different 
lids are plotted versus $(\epsilon-\epsilon_{\rm min})$.
The right panel shows \emph{i)} the same four data sets, now horizontally  
shifted in order to superimpose  the four lid values,  together with \emph{ii)} the LDOS for 
the  very small lid $\lambda=0.01$, also plotted as just described.
The LDOS   all admit the  exponential representation
\beq\label{expo}
 \varrho(\epsilon) = k \exp\left( \frac{\epsilon}{\alpha(\lambda)} \right).
 \eeq
 Their slope on a logarithmic vertical scale, $1/\alpha(\lambda)$,  systematically decreases with increasing 
lid, starting with an almost  vertical slope near $\epsilon_{\rm min}$. Note however that each straight line
covers many decades of variation of  the LDOS.
 Since the logarithm of the  latter quantity is  the  entropy,  the inverse slope
 $\alpha$ is nothing but the (micro-canonical) temperature. 
As one would expect, the latter  vanishes as the energy approaches the ground state
 energy, a limit enforced by $\lambda \rightarrow 0$.
 
\begin{figure}[h!]\label{dens_Gaussian}
 $
\begin{array}{cc}
\rotatebox{-90}{
\includegraphics[width=0.33\linewidth]{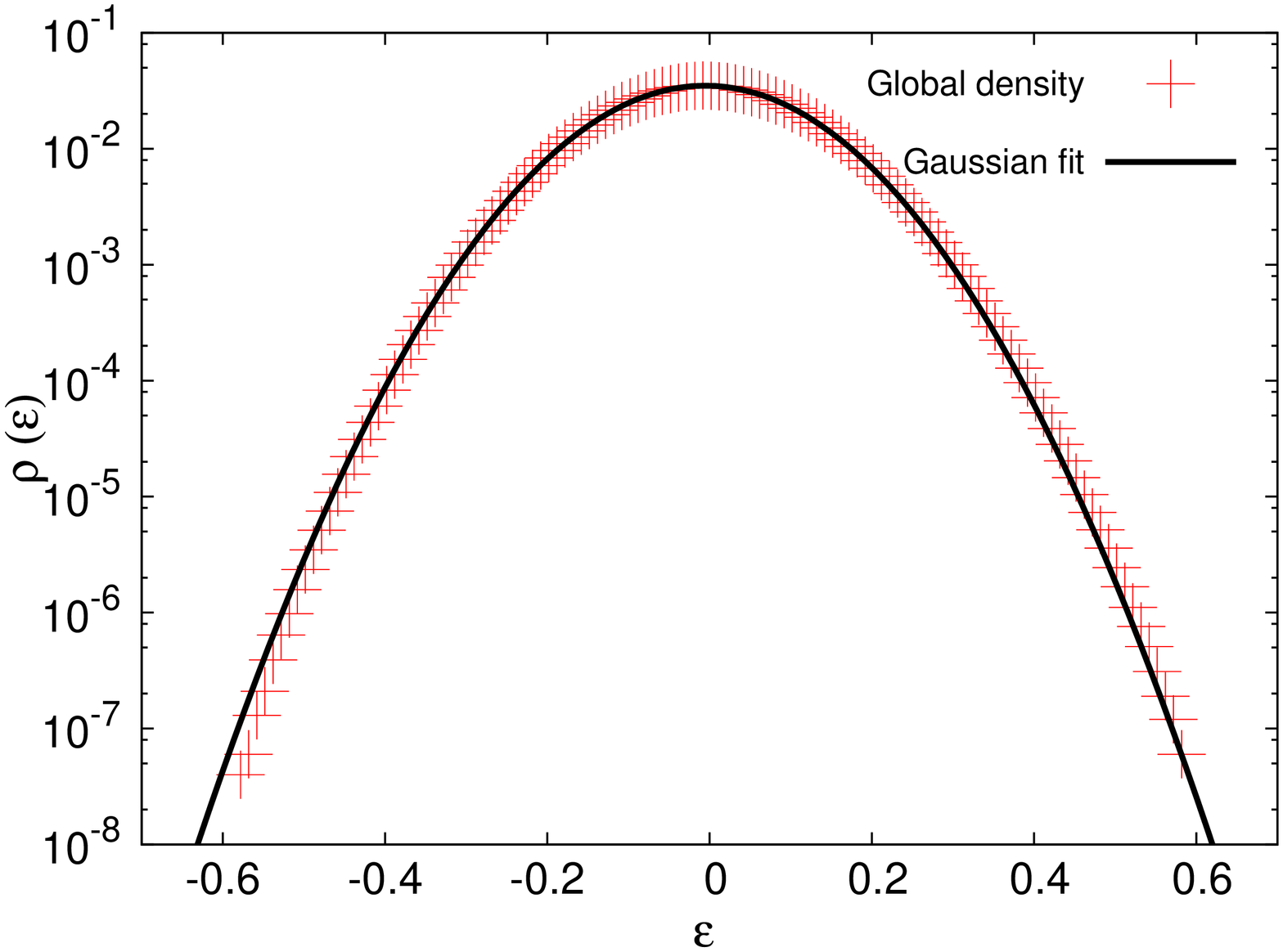}} & 
\rotatebox{-90}{\includegraphics[width=0.33\linewidth]{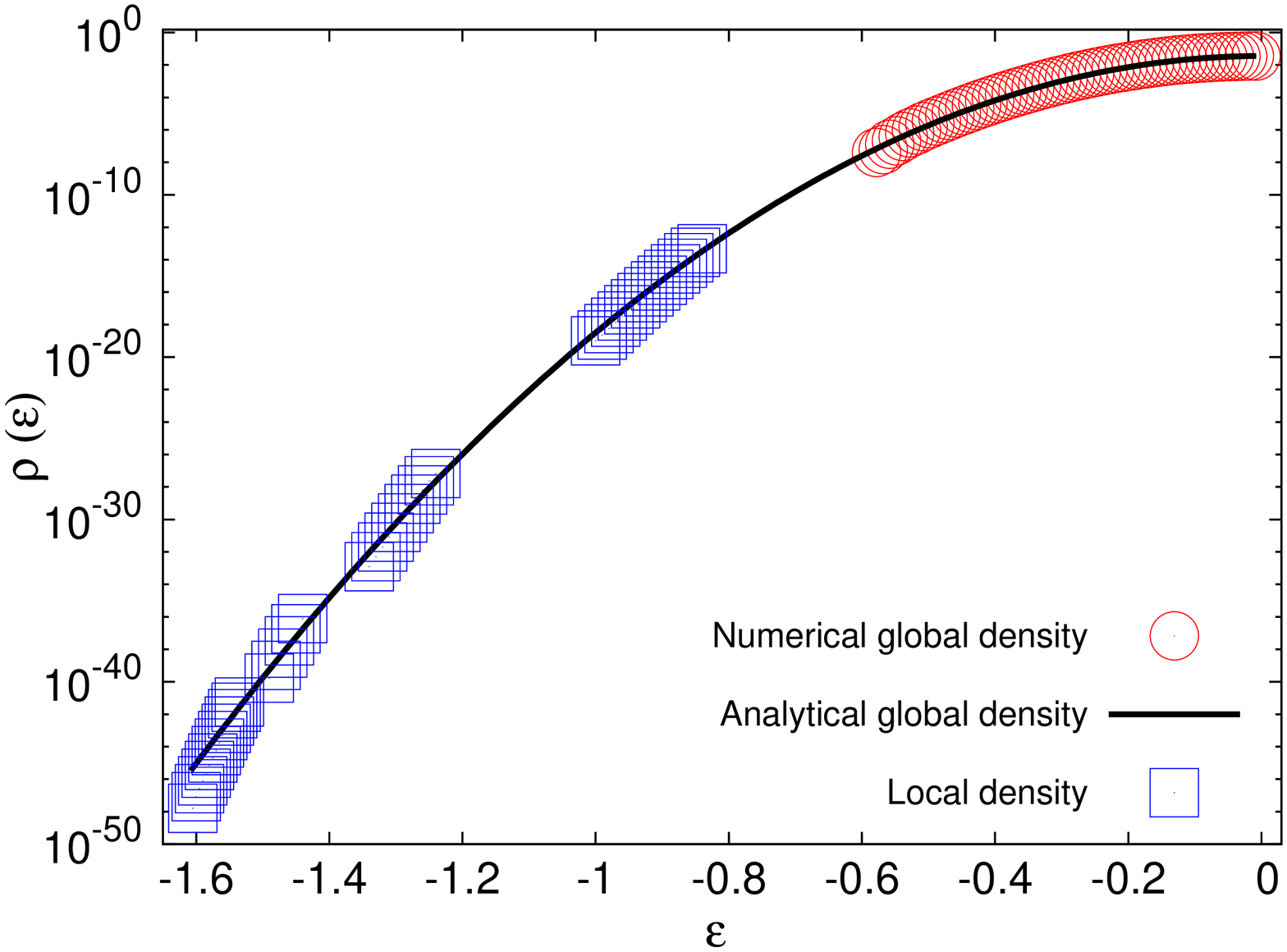} }
\end{array}
$ 
\caption{(Color online)
Left:  GDOS (plusses)  obtained by
running the entropic algorithm without lid restrictions and a Gaussian fit (line). 
 Right: The GDOS (line) obtained by extending the 
 range of the fit shown in the left panel. The LDOS
(squares) obtained 
for four different lids, $\lambda=0.1,0.2,0.4,08$. 
The numerical GDOS (circles) appears in the rightmost
corner of the figure. 
Note that the vertical axis spans $40$ 
order of magnitude.All data for a  3D lattice with $L=8$.}
\label{3Dgd} 
\end{figure}

 The left panel of  fig.~\ref{3Dgd}, shows the GDOS obtained  using $N_{h}=250$
 and without imposing any lid constraints. 
The global minimum  would  be  located on the abscissa 
at $\epsilon \approx -1.7$, i.e. far beyond the actual reach of the unrestricted entropic algorithm.
The  huge number of states present in the system simply prevents the algorithm from sampling a large 
fraction of the low energy states.
The line is a Gaussian fit, given by
 \beq\label{gaussian}
 \varrho(\epsilon) = k \exp\left( \frac{\epsilon}{b} -  \frac{\epsilon^2}{c} \right)\quad b=10;\quad c=0.025;\quad k=0.035.
 \eeq
The right panel shows the four different LDOS
vertically  rescaled  to make them approximately lie on the fitted GDOS curve.
These  data appear as straight line segments having
slope $1/\alpha(\lambda)$. The values of $\alpha$ for each segment are, in order of increasing lid value,
$\alpha =0.0049, 0.0058, 0.0071  $ and $0.0117$.
 A  Gaussian representation of the GDOS, as in the case of the REM model\cite{Derrida81, Bauke04},
 is  (by construction) inaccurate  near the minimum (and maximum) of the energy range,
and utterly fails  to reproduce the piecewise  lack of curvature characterizing  the  LDOS at low energies.

Confirming  and extending  arguments previously
 given in Refs,~\cite{Sibani93,Sibani94,Sibani99}, the exponential nature of the LDOS 
qualitatively explains why real 3D spin glasses lose their ability to thermally equilibrate 
 right below the critical temperature, see  Ref.\cite{Kenning09} and references therein. 
  The arguments given below are  valid in any glassy 
system with an exponential LDOS, and link landscape topography to aging behavior,
and in particular to  the origin of memory effects.

Expressed in terms
of the extensive energy $E=N\epsilon$, the Boltzmann equilibrium distribution
describing local equilibration in a valley is 
\beq\label{Boltz}
 P_{\rm Boltz}(E,T) = k \exp(E  \left[ \frac{1}{\alpha(\lambda)  N} -  \frac{1}{T} \right]).
 \eeq
 Here and in the following   $N$  
denotes  the number of spins  which are thermally active in the
valley considered, rather than the total number of spins in the system. Re-scaling the $\alpha$ 
values obtained for the LDOS reflects that the energy per spin $\epsilon$ is 
replaced  in Eq.~\eqref{Boltz} by  the extensive energy variable $E$. 
For local thermal equilibrium   within a valley to be possible, the sum in square
brackets must have negative sign, since this guarantees that  the `bottom' states of the valley
have the largest  probability. If the sign is instead  positive, the rim states have largest  probability.
The valley is then abandoned  with probability one and becomes  irrelevant to   the relaxation process.

 Consider now a system locally equilibrated at  $T>T_g$.  Decreasing the temperature, 
 even slightly, below $T_g$,   makes previously unaccessible valleys,
 namely those with  with $ \alpha N < T_g$    appear.   
  Hereby a  huge number of 
unexplored  configurations appears.
These are separated by energy barriers  corresponding to the internal structure of the `new' valleys. 
Effectively, a slight temperature decrease   quenches the system into  a 
local energy minimum,  and thereby  starts  the   aging process.
  Since the  value of $\alpha$ slowly decreases with energy,  some lower energy parts of the 
 energy landscape  remain  inaccessible  to the  aging  process for $T<\approx T_g$. 
 If the temperature is further decreased,   a new aging process starts in the yet unexplored part of the landscape.
Upon raising $T$  back to its previous value, the system recovers its previous
 local equilibrium state (a memory effect), and the effects of aging at the lower temperature are lost
 (a rejuvenation effect).
Rejuvenation and memory effects  are well  known in spin-glasses
   and other complex systems~\cite{Jonason98,Dupuis02,Josserand00,Berthier}.
 
 Our above arguments do not suffice to identify the value of $T_g$.  Given, however, 
 that the latter is known for the EA spin-glass,  $T_g=0.95$~\cite{Marinari}, we can,
 as a consistency check, estimate the number 
 of spins active in thermal  relaxation just below $T_g$.
 Using  the equilibrium average energy vs. temperature curve (not shown, but easily obtained using the density of states) 
  the equilibrium energy per spin at $T_g$ is found to be 
 $\langle \epsilon \rangle(T_g) \approx -1.05$. This value falls between the third and fourth energy intervals 
in which the LDOS was numerically  investigated. The average of the   corresponding two $\alpha$ values, 
$\alpha \approx 0.0094$ is therefore used to  estimate the logarithmic slope
 of the LDOS at that equilibrium energy. Using that $T_g$ is the temperature at which the manifold of  low-lying states 
 first appear, imposes  $ 0.0094 N= T_g  = 0.95$. This gives  $N \approx 101$ active  spins, which is slightly above a fifth of
  the $8^3 = 512$ spins present in the system.  

Let us finally consider a structural issue, i.e. the  topography of the energy landscape of the 3D
EA spin glass as it emerges   from the present investigations.
The exponential growth of the LDOS can be understood in the context of  a 
hierarchical picture of \emph{valleys within valleys}\cite{Sibani91,Sibani93,Sibani94}.
 The valley rooted at the ground state contains  sub-valleys whose number 
increases exponentially with energy. Each of these  has itself an exponentially 
growing number of internal sub-valleys, and so forth, down to 
a lower cutoff for the energy difference between top and bottom states
of a valley.
Since states at a given energy belong to a number of sub-valleys which 
increases exponentially with that energy, the LDOS itself grows  exponentially  in energy. 
A glance at Fig. \ref{3Dgd} indicates that the  picture is applicable for energies per spin below $\approx -0.7$.
At higher energies,
the logarithm of the LDOS has  non-negligible curvature, and the picture no longer  applies.  
\section{Discussion}
Unrestricted entropic algorithms~\cite{Ferrenberg88,Wang01,Lee93} can efficiently  
and  for a wide range of temperatures provide  thermal averages 
of relevant quantities in  several  models of physical
interest. If, however,   the  density of state varies   over e.g. forty  orders of magnitude
as is the case in the 3D EA model, 
 low-energy states  which only constitute a tiny fraction of the whole
are hard to sample efficiently and accurately.  
Using entropic algorithms  in lieu of  exhaustive enumeration\cite{Sibani93,Sibani94,Sibani99} 
 in connection with the lid algorithm  produces an accurate sampling of   the energy landscape 
of larger systems over a wider energy ranges than previously possible\cite{Sibani94,Klotz98a}.
As a consequence,  a  clear-cut difference between 2D and 3D landscape topography
becomes evident and  in 3D   quadratic fits of the logarithm of the density of states 
reveal   short-comings for energies near the ground state.
 
In 2D,  the  density of state has a  weak, nearly linear,  energy dependence,  except in a narrow region 
close to the ground state. The average energy is simply proportional to the temperature
and any thermal instability is ruled out.
The situation is completely different in 3D: 
while a parabola   describes   the energy dependence of the logarithm of the density of state 
very well over approximately $15$ orders of magnitude, this Gaussian  description   fails  near the 
ground state. Here,  the  logarithm of the  LDOS has  instead  a piecewise linear dependence
on the energy. Precisely this feature explains the inability of the spin-glass  
to equilibrate below $T_g$ and the instability of the thermalization dynamics to 
small temperature changes. From a structural point of view, the form of the LDOS in 3D  points to
a configuration space with nested valleys, and to an associated hierarchy of energy barriers,
properties which are widespread  in complex systems.
We thus  believe  that the method described in this work can  generally be of use  in mapping out in a computationally
efficient way geometrical properties of the low energy part of  the energy landscapes of complex systems. 
\section*{Acknowledgments}
PS would like to thank  Karl Heinz Hoffmann for his
hospitality and for an interesting discussion.

\begin{thebibliography}{99}
\bibitem{Wales03} David J. Wales. \textit{Energy Landscapes. With applications to Cluster,
 Biomolecules and Glasses}. Cambridge University Press, Cambridge, UK, (2003).
\bibitem{Stillinger83} F. H. Stillinger and T. A. Weber, Phys. Rev. A, 28, 2408-2416, (1983).
\bibitem{Ferrenberg88} A. M. Ferrenberg and R. H. Swendsen, Phys. Rev. Letters, 61, 2635--2638, (1988).
\bibitem{Wang01}   F. Wang and D. P. Landau, Phys. Rev. E, 64, 056101, (2001).
\bibitem{Lee93} J. Lee, Phys. Rev. 71, 211 (1993).
\bibitem{Sibani93} P. Sibani, C. Sch{\"{o}}n, P. Salamon and J.-O. Andersson. EPL, 22, 479-485, (1993).
\bibitem{Sibani94} P. Sibani and P. Schriver, Phys. Rev. B 49, 6667 (1993).
\bibitem{Schon00} J. C. Sch{\"{o}}n and P. Sibani, EPL 49, 196-202, (2000).
\bibitem{Schon98} J. C. Sch{\"{o}}n and P. Sibani, J. Phys. A, 31, 8165-8178, (1998).
\bibitem{Sibani99} P. Sibani, R. van der Pas and J. C. Sch{\"{o}}n, Computer Physics Communications, 116, 17-27, (1999).
\bibitem{Edwards75} S. F. Edwards and P. W. Anderson. J. Phys. F, 965-974, (1975).
\bibitem{Picco} M. Picco, F. Ricci-Tersenghi, and F. Ritort, Phys. Rev. B 63 174412 (2001).
\bibitem{Batt} R.N. Bhatt and A.P. Young, Phys. Rev. B 37, 5606 (1988).
\bibitem{Katzgraber06} H. G. Katzgraber, Mathias K\"{o}rnerand and A. P. Young, Phys. Rev. B 73, 224432 (2006).
\bibitem{Fischer91} K. H. Fischer and J. A. Hertz. \textit{Spin Glasses}. Cambridge University Press, (1991).
\bibitem{Parisi} M. Mezard, G Parisi, and M. Virasoro, \textit{Spin Glass Theory and Beyond}. World Scientific Lecture Notes in Physics Vol. 9 (World
Scientific, Singapore, 1987).
\bibitem{Klotz98a} T. Klotz, S. Schubert, and K. H. Hoffmann, Eur. Phys. J. B 2, 313-317 (1998).
\bibitem{Newman99b} M. E. Newman and G. T. Barkema. \textit{Monte Carlo Methods in Statistical Physics}. Oxford University Press (1999).
\bibitem{Marinari} E. Marinari, G. Parisi, and J.J. Ruiz-Lorenzo, Phys. Rev. B 58, 14852 (1998). 
\bibitem{Derrida81} B. Derrida, Phys. Rev. B 24, 2613 (1981).
\bibitem{Bauke04} H. Bauke and S. Mertens, Phys. Rev. E 70, 025102(R) (2004).
\bibitem{Kenning09} G. G. Kenning, J. Bowen, P. Sibani and G. F. Rodriguez, Phys. Rev. B (81), 014424 (2010)
\bibitem{Jonason98} K. Jonason, E. Vincent, J. Hammann, J. P. Bouchaud, and P. Nordblad, Phys. Rev. Lett. 81, 3243 (1998).
\bibitem{Dupuis02} V. Dupuis, E. Vincent, J.-P. Bouchaud, J. Hammann, A. Ito and H. Aruga Katori, Phys. Rev. B 64, 174204 (2002).
\bibitem{Josserand00} C. Josserand, A. V. Tkachenko, D. M. Mueth, and H. M. Jaeger, Phys. Rev. Lett. 85, 3632 (2000).
\bibitem{Berthier} L. Berthier and J.-P. Bouchaud, Phys. Rev. B 66, 054404 (2002).  
\bibitem{Sibani91} P. Sibani and K. H. Hoffmann, EPL 16, 423 (1991).
\end{thebibliography}

 \end{document}